\documentclass[10pt, conference, letterpaper]{IEEEtran}
\IEEEoverridecommandlockouts

\usepackage{cite}
\usepackage{amsthm}
\usepackage{amsmath,amssymb,amsfonts}
\usepackage[cmintegrals]{newtxmath}
\usepackage{algorithm}
\usepackage{algorithmicx}
\usepackage{enumerate}
\usepackage{algpseudocode}
\usepackage{caption}
\usepackage{bm}
\usepackage{subfig}
\usepackage{graphicx}
\usepackage{array}
\usepackage{textcomp}
\usepackage{xcolor}
\usepackage{url}
\usepackage{tabularx}

\makeatletter
\newcommand{\algmargin}{\the\ALG@thistlm}
\makeatother
\newlength{\whilewidth}
\algdef{SE}[parWHILE]{parWhile}{EndparWhile}[1]
  {\parbox[t]{\dimexpr\linewidth-\algmargin}{%
     \hangindent\whilewidth\strut\algorithmicwhile\ #1\ \algorithmicdo\strut}}{\algorithmicend\ \algorithmicwhile}%
\algnewcommand{\parState}[1]{\State%
  \parbox[t]{\dimexpr\linewidth-\algmargin}{\strut #1\strut}}

\newtheorem{definition}{Definition}
\newtheorem{lemma}{Lemma}
\def\BibTeX{{\rm B\kern-.05em{\sc i\kern-.025em b}\kern-.08em
    T\kern-.1667em\lower.7ex\hbox{E}\kern-.125emX}}

\begin{document}

\title{Performance Analysis for Correlated AoI and Energy Efficiency in Heterogeneous CR-IoT System
}

\author{\IEEEauthorblockN{Xiaoyu Hao\IEEEauthorrefmark{1}, Tao Yang\IEEEauthorrefmark{1}, Yulin Hu\IEEEauthorrefmark{2} and Bo Hu\IEEEauthorrefmark{1}}
\IEEEauthorblockA{\IEEEauthorrefmark{1}Research Center of Smart Networks and Systems, Dept. of Electronics Engineering, Fudan University, Shanghai, China.\\
\IEEEauthorrefmark{2}School of Electronic Information, Wuhan University, Wuhan, China.\\
Email: taoyang@fudan.edu.cn \vspace{-.2cm}}
}

\maketitle

\begin{abstract}
We consider a cognitive radio based Internet of Things (CR-IoT) system where the secondary IoT device (SD) accesses the licensed channel during the transmission vacancies of the primary IoT device (PD). We focus on the impact of the IoT devices' heterogeneous traffic pattern on the energy efficiency and on the age of information (AoI) performance of the SD. We first derive closed-form expressions of the energy efficiency and the average AoI, and subsequently explore their convexity and monotonicity to the transmit power. Following these characterizations, an optimal transmit power optimization algorithm (TPOA) is proposed for the SD to maximize the energy efficiency while maintaining the average AoI under a predefined threshold. Numerical results verify the different preferences of the SD toward different PD traffic patterns, and provides insights into the tradeoff between the energy efficiency~and~the~average~AoI.
\end{abstract}

\begin{IEEEkeywords}
cognitive radio based IoT, heterogeneous traffic, age of information, energy efficiency.
\end{IEEEkeywords}
\vspace{-.1cm}
\section{Introduction}
\IEEEPARstart{T}{he} Internet of Things (IoT) has become an important networking paradigm which enables massive connections among ubiquitous physical objects. To address the spectrum scarcity caused by large-scale IoT device access and low spectrum efficiency of static spectrum allocation, a promising solution is to apply the cognitive radio (CR) technology to the IoT, which is well-known as CR-IoT \cite{CRIoT2017Survey}. The CR technology enables the IoT device without dedicated spectrum to work as a secondary user (SU) and accesses the licensed channel of the nearby legitimate IoT device, i.e. the primary users (PU), without causing performance degradation to primary service. There are mainly three CR spectrum sharing strategies, including underlay, overlay, and interweave schemes.
% \cite{2017CRSurvey}
Due to the ease of implementation, the interweave scheme is more preferred and extensively adopted in the CR-IoT system, where the SU first monitors the status of the licensed channel and accesses the channel only when it is not occupied by the PU. 

\par
Massive emerging IoT services could be promoted by applying the CR technology, such as smart cities, pollution control, wildfire monitoring and smart agriculture, etc. In these scenes, the wireless connected, battery-operated IoT devices are deployed to monitor certain time-critical physical processes, while there are two common concerns in the design of such systems. One is the battery lifetime, since replacing batteries usually incurs high cost, while the CR functionalities, e.g., channel sensing and switching, are energy consuming. Thus, the CR scheme should be carefully designed to maximize the energy efficiency. The second one is the information freshness, since outdated state information loses value and may even cause severe accidents. The freshness of information can be evaluated by a new concept, i.e. age of information (AoI), which is defined as the time elapsed since the most recent received update was generated at the source~\cite{2012AoI}. 

\par
The AoI has been investigated as an important performance metric in the cognitive radio networks (CRNs) \cite{2019EHCR,2020CRCollision,CRAoI2017,2019IoTJUorO}. The authors in \cite{2019EHCR} investigate the optimal sensing and update scheme of an energy harvesting CR-based sensor for AoI minimization, taking into consideration the partially observability of the state of the PU. Instead of considering slotted transmission and strict slot synchronization between the PU and the SU, the work in \cite{2020CRCollision} focuses on the unsynchronized case and formulates the scheduling policy design problem of the SU for the average AoI minimization as a Markov decision process (MDP) with a collision constraint. The authors in \cite{CRAoI2017} consider an interweave-based cognitive wireless sensor network, and propose a joint framing and scheduling policy   optimizing the  energy efficiency under strict expected AoI constraints. In \cite{2019IoTJUorO}, the underlay scheme and the overlay scheme are compared with each other with respect to the average peak AoI of both the PU and the SU under standard ARQ. 

\par
However, the above existing studies haven't investigated the impact of heterogeneous traffic patterns of IoT devices on the energy efficiency and on the AoI performance in the CR-IoT system. The primary IoT device (PD) and the secondary IoT device (SD) may have distinct traffics, e.g. the packet generation rate and data size. The energy efficiency and the AoI of the SD may be quite different under different PD traffic patterns, since in the interweave mode, the SD can only transmit when the PD is idle, and the secondary transmission may be frequently interrupted by the arrival of primary traffic. Hence, in this work we are motivated to focus on the effect of heterogeneous traffic on the energy efficiency and AoI of the SD in the CR-IoT system. The main contributions of this article are summarized as follows:
\begin{itemize}
  \item 
With the consideration of the randomness of spectrum access in the considered CR-IoT system, we derive the closed-form expressions of the energy efficiency and average AoI of the SD. In particular, we show that the average AoI tends to infinity under two extreme cases, which implies the AoI performance is closely related to the specific traffic patterns.

  \item 
We explore how the energy efficiency and the average AoI evolves with the transmit power, and prove the convexity of the average AoI as well as the convexity and monotonicity of the energy consumption w.r.t the required transmission time. With the above properties, we propose an optimal transmit power optimization algorithm for the SD to maximize its energy efficiency while maintaining the average AoI under the predefined threshold. 
\end{itemize}
\par
The remainder of this paper is organized as follows. We introduce the system model in Section II and derive the closed-form expression of the energy efficiency and average AoI in Section III. An energy-efficient, AoI aware power optimization scheme is proposed in Section IV. In Section V, numerical results are reported with discussions.

\section{System Model}
We consider a CR-IoT system, where the SD with no dedicated spectrum performs a certain remote monitoring task and opportunistically accesses the licensed channel legitimate to the PD to update the monitored status information to a secondary access point (SAP). From the SD's point of view, the availability of a licensed channel can be modelled as two states, i.e. IDLE and BUSY, which correspond to the two cases where the PD is or isn't utilizing the channel, respectively. We assume that the state transition of the channel follows a two-state continuous-time Markov Chain (CTMC), which is a reasonable and widely-adopted assumption \cite{CTMC2018,CTMC2020}. Let $u$ and $v$ be the transition rates from IDLE to BUSY and from BUSY to IDLE, which jointly represent the traffic pattern of the PD. Then, the continuous IDLE and BUSY periods denoted by $T^I$ and $T^B$ are independent and exponentially distributed random variables with mean value $1/u$ and $1/v$, respectively. 
\par
The SD monitors a physical process which randomly generates status updates of size $D$ bits according to a Poisson process of rate $\lambda$, and thus the traffic pattern of the SD is modeled by the packet size and generation rate. The terms status update and packet are used interchangeably throughout this paper. We assume that the SD can simultaneously handle or hold only one packet, and thus the packets that arrive during transmission will be discarded. Besides, for information freshness, newly-generated packets will take the place of the old one when the SD is waiting for transmission opportunities. We assume the SD transmits at the Shannon capacity $C=B\log_2{\left(1+\frac{P^T}{{N_0}B}\right)}$, where $P^T$ is the transmission power and $N_0$ is the equivalent noise power per unit bandwidth at the receiver in consideration of the channel effect\footnote{For analysis simplicity, the equivalent noise power only accounts for the path loss, which can be seen as a fixed value. However, the analysis in this paper, more specifically, the derivation of energy efficiency and expected AoI in Section III, can be easily extended to the block fading case.}.
During transmission, if the PD reclaims the channel, the SD has to quit the current transmission, hand over the channel to the PD and wait for the next IDLE period to restart the transmission. 

\subsection{Age of Information}
\begin{figure}[!t]
\centering
\includegraphics[width=2.7 in, trim= 0 35 0 20]{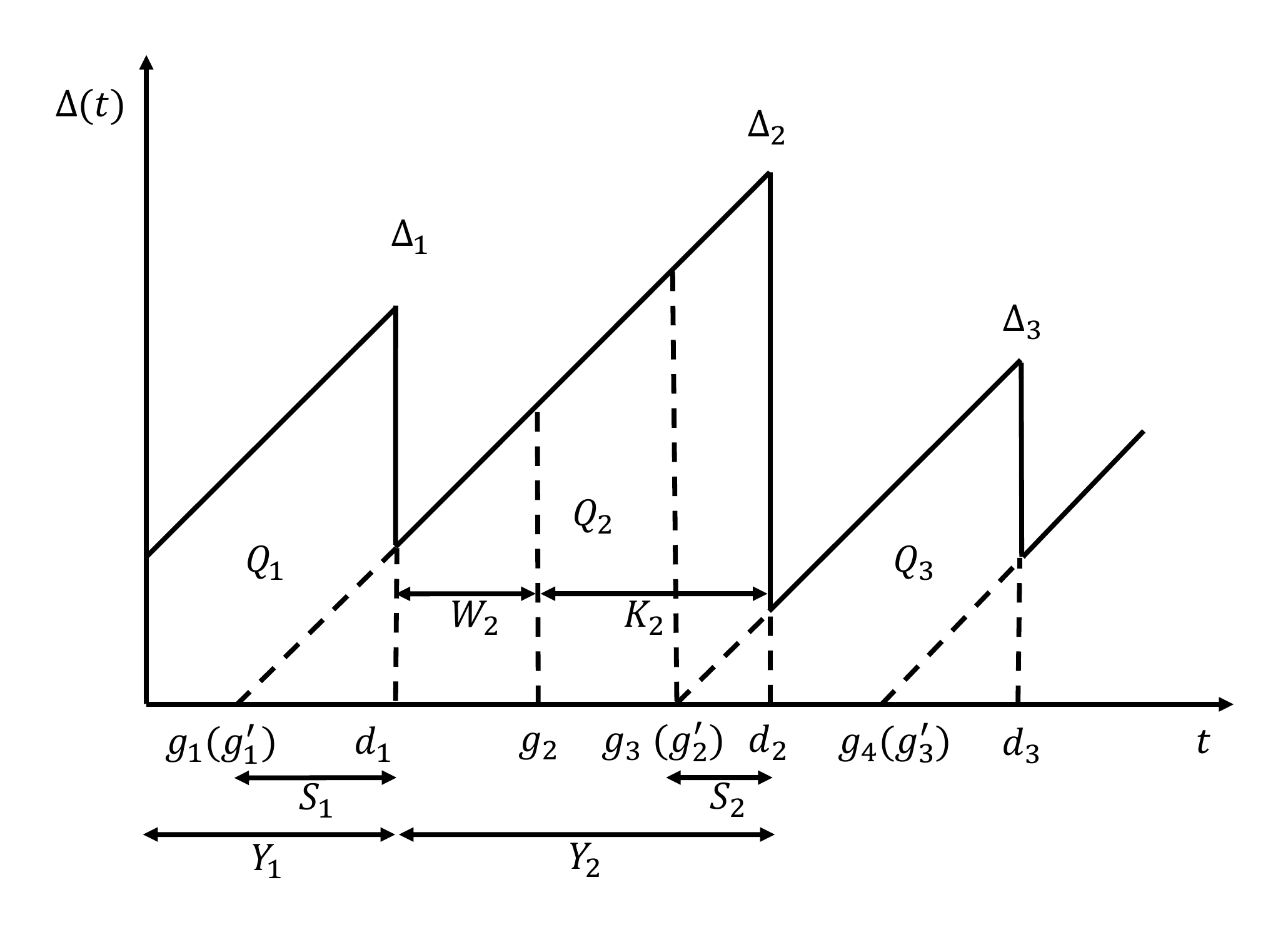}
\captionsetup{font={small}}
\caption{A sample path of AoI and the corresponding intervals.}
\label{AoISamplePath}
\vspace{-.53cm}
\end{figure}

Denote the generation time of the $i$-th packet by $g_i$. Note that not every generated packet will be finally received at the SAP, since packets generated during the transmission will be discarded, and packets to be transmitted will be replaced by the newly-generated ones. Thus we denote by $g_i^\prime$ and $d_i$ the generation time and the departure time of the $i$-th successfully transmitted packet. At time instant $t$, we define the index of the most recently received packet at the SAP as $N_t=\max{\{i\mid d_i \leq t\}}$. Then, the instantaneous AoI at $t$ is defined as follows \cite{2012AoI}.
\vspace{-.2cm}
\begin{definition} \
  \label{def:AoI}
  $\!\!\!$An~instantaneous~AoI~at~time~point~$t$~is~defined~as
  \vspace{-.35cm}
  \begin{equation}\vspace{-.2cm}
    \label{AoIdef}
    \Delta (t) = t - g_{N_t}.
  \end{equation}
\end{definition}
A sample path of AoI is illustrated in Fig. \ref{AoISamplePath}. To calculate the average AoI, we define the interval between the $i$-th and $(i-1)$-th departures as
\vspace{-.2cm}
\begin{equation}\vspace{-.1cm}
  \label{Ydef}
  Y_i = d_i - d_{i-1},
\end{equation}
and define the service time of the $i$-th successful received packet as
\vspace{-.1cm}
\begin{equation}\vspace{-.1cm}
  \label{Sdef}
  S_i = d_i - g_i^\prime.
\end{equation}
$Y_i$ in (\ref{Ydef}) can be divided into two intervals: $W_i$ and $K_i$. $W_i$ is defined as the time elapsed since the last departure $d_{i-1}$ until a new packet is generated, which can be expressed as
\vspace{-.1cm}
\begin{equation}\vspace{-.1cm}
  \label{Wdef}
  W_i = \min{\{g_i \left.\right| g_i\geq d_{i-1} \}} - d_{i-1},
\end{equation}
The other interval $K_i$ is defined as the time elapsed since the first packet generation after the last departure until the next successful reception $d_i$, which can be expressed as
\vspace{-.1cm}
\begin{equation} \vspace{-.1cm}
  \label{Kdef}
  K_i = d_i - \min{\{g_i \left.\right| g_i\geq d_{i-1}\}}.
\end{equation} 
Following the above definitions, the average AoI can be calculated based on the polygon area $Q_i$ depicted in Fig.\ref{AoISamplePath}:
\vspace{-.1cm}
\begin{equation}\vspace{-.2cm}
  \label{AoIArea}
  \bar{\Delta}
  =\lim_{t \to \infty}{\frac{N_t}{t} \frac{1}{N_t} \sum_{i=1}^{N_t}{Q_i}}
  =\frac{\mathbb{E} [Q_i]}{\mathbb{E} [Y_i]},
\end{equation}
where 
\vspace{-.2cm}
\begin{equation}
  \label{QArea}
  Q_i 
  = \frac{(S_{i-1}+Y_i)^2}{2} - \frac{S_{i-1}^2}{2} 
  = \frac{Y_i^2}{2} + S_{i-1}Y_i.
\end{equation}
Then, we have the following lemma representing the relationship between 
the service time and the inter-departure time. % has the following .

\vspace{-.1cm}
\begin{lemma}
  \label{lemma_S&Y}
  $S_{i-1}$ is independent of $Y_i$.
\end{lemma}
\begin{IEEEproof}
Since $Y_i=W_i+K_i$, $\mathbb{E}\left[ S_{i-1} Y_i \right]=\mathbb{E}\left[S_{i-1}\right]\mathbb{E}\left[Y_i\right]$ holds if $S_{i-1}$ is independent of both $W_i$ and $K_i$. The packet generation process, which follows an Poisson process, is independent of the state transition of the channel as well as the packet transmission. $W_i$ is the waiting time from the $i-1$-th departure to the generation of the first packet after $d_{i-1}$, and thus follows an exponential distribution and has the memoryless property. Therefore, $W_i$ is independent of the events happened before $d_i$ and thus $W_i$ is independent of $S_{i-1}$. $K_i$ is the time elapsed from the generation of a first packet after $d_{i-1}$ to the next departure $d_i$, and thus it depends on the channel state transition process and the packet generation process. Note that the former has the Markovian property and the latter has the memoryless property, so $K_i$ is also independent of the events happened before the packet generation. Therefore, $K_i$ is independent of $S_{i-1}$, and thus Lemma~\ref{lemma_S&Y} is proved.  
\end{IEEEproof}

According to Lemma~\ref{lemma_S&Y}, the expression of the average AoI can be simplified as
\vspace{-.1cm}
\begin{equation}\vspace{-.1cm}
  \label{AoISimplified}
  \bar{\Delta}
  \!=\!\mathbb{E}[S_{i-1}] \!+\! \frac{\mathbb{E}[Y_i^2]}{2\mathbb{E}[Y_i]}
  \!=\!\mathbb{E}[S_{i-1}] \!+\! \frac{\!\mathbb{E}[W_i^2]\!+\!2\mathbb{E}[W_i K_i]\!+\!\mathbb{E}[K_i^2]\!}{2(\mathbb{E}[W_i]\!+\!\mathbb{E}[K_i])}.
\end{equation}
\par
Note that the sequences $\{W_1,W_2,\cdots\}$, $\{K_1,K_2,\cdots\}$, $\{Y_1,Y_2,\cdots\}$ and $\{S_1,S_2,\cdots\}$ form i.i.d processes, which allows us to drop the subscript index of $W_i$, $K_i$, $Y_i$ and $S_{i-1}$ in (\ref{AoISimplified}). As a result, the average AoI can be further~reformulated~as
\vspace{-.1cm}
\begin{equation}\vspace{-.1cm}
  \label{AoIDefFinal}
  \bar{\Delta}
  = \mathbb{E}[S] + \frac{\mathbb{E}[Y^2]}{2\mathbb{E}[Y]}
  = \mathbb{E}[S] + \frac{\mathbb{E}[W^2]+2\mathbb{E}[WK]+\mathbb{E}[K^2]}{2(\mathbb{E}[W]+\mathbb{E}[K])},
\end{equation}
where $W$ represents the waiting time from a successful reception to the generation of a new status update, $K$ represents the time elapsed from the first packet generation after a successful reception to the next successful reception, and $S$ represents the service time of a successfully transmitted packet.

\subsection{Energy Efficiency}
Energy efficiency in this work is defined as the average number of successfully transmitted bits per unit energy consumption. We assume that the power consumption of the SD consists of three parts \cite{2016CRPowerConsumption}: the packet transmit power $P^T$, the static circuit power $P^C$ and the spectrum sensing power $P^S$. We denote by $E^T$, $E^S$ and $E^C$ the average energy consumption of packet transmission, spectrum sensing and static circuit operation between two consecutive departures, respectively. Besides, we define the corresponding average power consumption time as $T^T$, $T^S$ and $T^C$. Based on the above definitions, the energy efficiency can be expressed as
\begin{equation}
  \label{EEdef}
  {EE}
  = \frac{D}{E^T+E^S+E^C}
  = \frac{D}{P^T T^T + P^S T^S + P^C T^C}.
\end{equation}

\par
So far, we have introduced the general models of AoI and the energy efficiency. In the following section, we derive the explicit expressions of them.

\section{Correlation Analysis between energy efficiency and average AoI}
As for the energy efficiency, the static circuit power consumption lasts for the whole interval between two departures, so $T^C=\mathbb{E}[Y]$. After a successful reception, the SD starts to sense the allocated channel when a new packet is generated, and keeps monitoring the channel state until a packet is successfully received, i.e. $T^S=\mathbb{E}[K]$. In this way, the SD can release the channel immediately on arrival of the PD's traffic and access the channel when the next transmission opportunity comes. As a result, to evaluate the energy efficiency, we need to derive $\mathbb{E}[Y]$, $\mathbb{E}[K]$ and $T^T$.
As for evaluating the average AoI, according to \eqref{AoIDefFinal}, the main difficulty is to derive the terms $\mathbb{E}[S]$, $\mathbb{E}[Y]$ and $\mathbb{E}[Y^2]$. 
Hence, in the following we first derive the expressions of $\mathbb{E}[Y]$, $\mathbb{E}[Y^2]$ and $T^T$ in Section III-A as well as $\mathbb{E}[S]$ in Section III-B. %The term of $\mathbb{E}[S]$ is specifically derived in Section III-B.

\subsection{Characterization on $\mathbb{E}[Y]$, $\mathbb{E}[Y^2]$ and $T^T$}
The packet generation process follows a Poisson process of rate $\lambda$, which is a memoryless process. Thus, the elapsed time for generating a packet from a certain instant follows an exponential distribution of parameter $\lambda$. Then $W$, which is the time elapsed from the last departure to the generation of a new packet, also follows the exponential distribution of parameter $\lambda$. So the probability density function (PDF) denoted by $f_W(t)$ is $f_W(t) = \lambda e^{-\lambda t}$, and we have $\mathbb{E}[W]=1/{\lambda}$, $\mathbb{E}[W^2]=2/{{\lambda}^2}$.

\par
To derive the first and second moment of $K$, note that $K$ evolves differently for different channel states at the initial moment. Specifically, if the channel is in IDLE state when a new packet is generated, the SD can immediately start the transmission. On the other hand, if the channel is in BUSY state, the SD has to wait until the PD finishes its transmission. We denote by $I_K$ and $B_K$ the events that the channel is in IDLE or BUSY state at the initial moment of $K$, respectively. Then the expectation of $K$ can be expressed as
\begin{equation}
  \label{conditionalK}
  \mathbb{E}[K] = 
  \mathrm{Pr}\{I_K\} \mathbb{E}[K | {I_K}] + 
  \mathrm{Pr}\{B_K\} \mathbb{E}[K | {B_K}],
\end{equation}
and the corresponding expectation of $K^2$ is given by
\begin{equation}
  \label{conditionalK2}
  \mathbb{E}[K^2] = 
  \mathrm{Pr}\{ I_K \} \mathbb{E} [{K^2}| {I_K} ] + 
  \mathrm{Pr}\{ B_K \} \mathbb{E} [{K^2}| {B_K} ].
\end{equation}
\par
The two probability terms in (\ref{conditionalK}) and (\ref{conditionalK2}) can be calculated from the transition probabilities of the channel CTMC. The channel must be in the IDLE state at the last departure, so the probability of $I_K$ and $B_K$ equals the probability that the CTMC transfers from IDLE to IDLE and BUSY after a period of $W$, respectively. Denote by $P_{II}(t)$ and $P_{IB}(t)$ the transition probability from IDLE to IDLE and from IDLE to BUSY after time $t$, which are given by \cite{medhi2002stochastic}
\begin{equation}
  \label{P_II}
  P_{II}(t) = \frac{v}{u+v}+\frac{u}{u+v} e^{-(u+v)t},
\end{equation}
\begin{equation}
  \label{P_IB}
  P_{IB}(t) = \frac{u}{u+v}-\frac{u}{u+v} e^{-(u+v)t}.
\end{equation}
\par
According to the transition probabilities and the PDF of $W$, $\mathrm{Pr}\{I_K\}$ and $\mathrm{Pr}\{B_K\}$ can be obtained as follows:
\begin{equation}
  \label{conditionalK_I}
  \mathrm{Pr} \left\{ I_K \right\} = 
  \int_0^{+\infty} P_{II}(t) \lambda e^{-\lambda t} dt =
  \frac{v+\lambda}{u+v+\lambda},
\end{equation}
\begin{equation}
  \label{conditionalK_B}
  \mathrm{Pr} \left\{ B_K \right\} = 
  \int_0^{+\infty} P_{IB}(t) \lambda e^{-\lambda t} dt =
  \frac{u}{u+v+\lambda}.
\end{equation}
\par
Since the IDLE period $T^I$ and the BUSY period $T^B$ are both exponential random variables, they also have memoryless property. Inspired by \cite{gu2019timely,2019IoTJUorO}, the two conditional expectation terms $\mathbb{E}[K|{I_K}]$ and $\mathbb{E}[K|{B_K}]$ in (\ref{conditionalK}) can be evaluated in a recursive manner. Denote by $t^P$ the time required for the SD to transmit an entire packet at the Shannon rate, which is calculated as $t^P=D/C$. As for $\mathbb{E}[K|{I_K}]$, the channel is IDLE at the beginning of $K$, and thus the SD can transmit the generated packet immediately. There are two cases. The first one is that $t^P$ is shorter than the IDLE period $T^I$. In this case the SD finishes its transmission without being interrupted by the PD traffic. The second one is that the channel state transfers from IDLE to BUSY during the SD transmission. In this case, the SD has to quit its current transmission and wait until the channel turns to IDLE again. With the memoryless property of $T^B$, the time elapsed from the instant that channel state transfers to BUSY to the next departure is equal to that the channel is occupied at the beginning of $K$. With the analysis above, $\mathbb{E}[K|{I_K}]$ can be evaluated as follows 
\vspace{-.1cm}
\begin{equation}\vspace{-.1cm}
  \label{K_I_recursive}
  \mathbb{E}[K|{I_K}] = 
  t^P \int_{t^P}^{+\infty}{ue^{-ut}dt}+
  \int_0^{t^P}\left({t+\mathbb{E}[K|{B_K}]}\right) {ue^{-ut}dt}.
\end{equation} 
The first term on the right side of (\ref{K_I_recursive}) corresponds the first case, where $\mathbb{E}[K|{I_K}]$ equals the required transmission time for one packet. The second term represents that $\mathbb{E}[K|{I_K}]$ is composed of the already transmitted time before interruption and the expected elapsed time before the next departure when the channel happens to be BUSY at the packet generation moment.
In the same way, $\mathbb{E}[K|{B_K}]$ can be expressed as
\vspace{-.1cm}
\begin{equation}\vspace{-.1cm}
  \label{K_B_recursive}
  \mathbb{E}[K|{B_K}] = 
  \int_0^{t^P}\left({t+\mathbb{E}[K|{I_K}]}\right) {ve^{-vt}dt}.
\end{equation} 
which is composed of the waiting time before the next transmission opportunity and the expected time before a successful reception when the channel is IDLE at the initial moment of $K$.
By jointly considering (\ref{K_I_recursive}) and (\ref{K_B_recursive}), we can obtain the expressions of $\mathbb{E}[K\left|{I_K}\right.]$ and $\mathbb{E}[K\left|{B_K}\right.]$. Substitute the results along with (\ref{conditionalK_I}) and (\ref{conditionalK_B}) to (\ref{conditionalK}), and $\mathbb{E}[K]$ is expressed as
\vspace{-.1cm}
\begin{equation} \vspace{-.1cm}
  \label{K_final}
  \mathbb{E}[K] = 
  h e^{ut^P} +
  \frac{u}{v(u+v+\lambda)} - h,
\end{equation}
where $h = 1/u + 1/v$,
and $\mathbb{E}[Y]$ can be finally obtained as
\vspace{-.1cm}
\begin{equation}\vspace{-.1cm}
  \label{Y_final}
  \mathbb{E}[Y]  = \frac{1}{\lambda} + h e^{ut^P} + \frac{u}{v(u+v+\lambda)} - h.
\end{equation}
\par
Since all interruptions happen with identical probability, which only depends on the required packet transmission time of the SD and the channel CTMC, the number of transmission interruptions follows the Geometric distribution of probability $p^I = \mathrm{Pr} \left\{T^I<t^P\right\}= 1 - e^{-ut^P}$. 
We define $t^I$ as the expected time spent for transmission before interruption, given by % and it can be calculated as
\vspace{-.1cm}
\begin{equation}\vspace{-.1cm}
  \label{TimeBeforeInterruption}
  t^I = 
  \frac{\int_0^{t^P}{tue^{-ut}}}{p^I} =
  \frac{1}{u} - \frac{e^{-ut^P}}{1-e^{-ut^P}} t^P.
\end{equation}
So the expected transmission time spent between two departures can be evaluated as the sum of the interrupted transmissions and one successful transmission, given by.
\begin{equation}
  \label{TT}
  T^T = \sum_{n=0}^{+\infty}{{p^I}^n\left(1-p^I\right)} n t^I + t^P = \frac{e^{ut^P}-1}{u}.
\end{equation}
\par
Next we turn to the two conditional expectations of $K^2$. Denote by $f_{K|I_K}(t)$ and $f_{K|B_K}(t)$ the PDF of $K$ conditioned on $I_K$ and $B_K$, respectively. Similar to the analysis on $\mathbb{E}[K|{I_K}]$, $\mathbb{E}[{K^2} | {I_K}]$ can be derived based on (\ref{K_I_recursive}) as follows:\vspace{-.1cm}
\begin{align}
  \label{K2_I_recursive}
  \mathbb{E}\left[K^2\left|{I_K}\right.\right] = {} & \underbrace{{t^P}^2 \int_{t^P}^{+\infty}{ue^{-ut}dt}}_{S_1} \notag \\
  {} & + \underbrace{{\int_0^{t^P}{\int_0^{+\infty}{(t+s)^2 {f_{K|I_K}(s)} u e^{-ut} ds dt}}}}_{S_2}.
\end{align}
The terms $S_1$ and $S_2$ correspond to the first and second terms on the right side of (\ref{K_I_recursive}), respectively. $S_2$ can be further transformed into
\vspace{-.1cm}
\begin{equation}\vspace{-.1cm}
  \label{S2}
  S_2 = \int_0^{t^P} {
    \left(
    {
    t^2 + 
    2t \mathbb{E} [{K}| {B_K}] +
    \mathbb{E} [{K^2}| {B_K}]
    }
    \right)
    ue^{-ut}dt
    }.
\end{equation}
Similarly, based on (\ref{K_B_recursive}). the expectation of $K^2$ conditioned on $B_K$ can be expressed as
\vspace{-.1cm}
\begin{equation}\vspace{-.1cm}
  \label{K2_B_recursive}
  \mathbb{E}[K^2|{B_K}] 
  = \int_0^{t^P} {
  \left(
  {
  t^2 + 
  2t \mathbb{E} [{K}| {I_K}] +
  \mathbb{E} [{K^2}| {I_K}]
  }
  \right)
  ve^{-vt}dt
  }.
\end{equation}
With the derived results of $\mathbb{E}[K|{I_K}]$ and $\mathbb{E}[K|{B_K}]$, $\mathbb{E}[K^2|{I_K}]$ and $\mathbb{E}[K^2|{B_K}]$ can be solved from (\ref{K2_I_recursive}), (\ref{S2}) and (\ref{K2_B_recursive}). We omit the two expressions here and directly give the result of $\mathbb{E}[K]$ as follows
\vspace{-.1cm}
\begin{align}
  \label{K2_final}
  \mathbb{E}[K^2] = {} & \frac{2}{uv} - \frac{2}{v(u+v+\lambda)} + 2h^2 e^{2ut^P} \notag \\
  {} & + 2\left[ 
  h \left( \frac{u}{v(u+v+\lambda)} - h - t^P \right) - \frac{1}{uv}
  \right]e^{ut^P}.
\end{align}
\par
Now we derive the expression of $\mathbb{E}[WK]$. We can find that $W$ and $K$ are independent of each other when conditioned on $I_K$ and $B_K$. Then, we have
\begin{align}
  \label{WK}
  \mathbb{E}[WK] = {} &
  \mathbb{E}[ WK | I_K ] \mathrm{Pr} \{I_K\} + 
  \mathbb{E}[ WK | B_K ] \mathrm{Pr} \{B_K\} \notag \\
  = {} &
  \mathbb{E}[ W | I_K ] \mathbb{E}[ K | I_K ] \mathrm{Pr} \{I_K\} +
  \mathbb{E}[ W | B_K ] \mathbb{E}[ K | B_K ] \mathrm{Pr} \{B_K\} 
\end{align}
For the term $\mathbb{E}[ WK | I_K ]$, denote by $f_{W|I_K}(t)$ the PDF of $W$ conditioned on event $I_K$, then it can be calculated as
\begin{align}
  \label{W_I_K}
  \mathbb{E}[ W | I_K ] = {} &
  \int_0^{+\infty}{f_{W | I_K }(t)tdt} %\notag \\
  =  \int_0^{+\infty} { \frac{P_{II}(t)f_W(t)}{\mathrm{Pr}\{ I_K \}}tdt} \notag \\
  = {}& \left( \frac{1}{u\lambda} + \frac{\lambda}{v(u+v+\lambda)^2} \right) \left( e^{ut^P} - 1 \right).
\end{align}
Similarly, $\mathbb{E}[ W | B_K ]$ is given by
\begin{equation}
  \label{W_B_K}
  \mathbb{E}[ W | B_K ] = \left( \frac{1}{v\lambda} - \frac{\lambda}{v(u+v+\lambda)^2} \right) 
  \left( e^{ut^P} - \frac{v}{u+v} \right).
\end{equation}
By substituting (\ref{conditionalK_I}), (\ref{conditionalK_B}), (\ref{W_I_K}) and (\ref{W_B_K}) into (\ref{WK}), $\mathbb{E}[WK]$ is finally given as follows
\begin{equation}
  \label{WK_final}
  \mathbb{E}[WK] = \frac{1}{\lambda}\left[ he^{ut^P} + \frac{u}{v(u+v+\lambda)} - h \right]
  + \frac{u}{v(u+v+\lambda)^2}.
\end{equation}
Finally, the expression of $\mathbb{E}[ Y^2 ]$ can be obtained as
\begin{equation}
  \label{Y2_final}
  \mathbb{E}[ Y^2 ] =  2\left\{h^2e^{2ut^P} + \left[ h\left( g - t^P \right) - \frac{1}{uv} \right]e^{ut^P} + \frac{g}{\lambda} + q \right\}
\end{equation}
where $g = \frac{1}{\lambda} + \frac{u}{v(u+v+\lambda)} - h$ and $q = \frac{1}{uv} - \frac{v+\lambda}{v(u+v+\lambda)^2}$, 
and the $\mathbb{E}[Y^2]/2\mathbb{E}[Y]$ term in (\ref{AoIDefFinal}) can be expressed as
\begin{equation}
  \label{Y2_2Y_final}
  \frac{\mathbb{E}[ Y^2 ]}{2\mathbb{E}[Y]} = \frac{h^2 e^{2ut^P} + \left[ h(g-t^P) - \frac{1}{uv} \right]e^{ut^P} + \frac{g}{\lambda} + q}
  {he^{ut^P} + g}.
\end{equation}

\subsection{Characterization on $\mathbb{E}[S]$}
Similar to $K$, to derive $\mathbb{E}[S]$, we also need to distinguish different channel states at the initial moment. Define $I_S$ and $B_S$ as the event that the channel is IDLE and BUSY at the initial moment of $S$. Then the expectation of $S$ follows
\vspace{-.1cm}
\begin{equation}\vspace{-.1cm}
  \label{S}
  \mathbb{E}[S] = \mathbb{E}[ S | I_S ] \mathrm{Pr} \{ I_S \} + 
  \mathbb{E}[ S | B_S ] \mathrm{Pr}\{ B_S \}.
\end{equation}
Both $I_S$ and $B_S$ imply the event denoted by $\Phi_S$ that packet is finally successfully received at the destination without being replaced by a new one. In this sense, we let $I_G$ and $B_G$ represent the events that the channel is IDLE and BUSY when a packet is generated, and then $I_S$ and $B_S$ represent the events $I_G$ and $B_G$ conditioned on $\Phi_S$, respectively. Therefore, the probability of event $I_S$ and $B_S$ can be calculated as
\vspace{-.1cm}
\begin{equation}\vspace{-.1cm}
  \label{Pr_I_S}
  \mathrm{Pr} \{ I_S \}
  = \frac{\mathrm{Pr}\{ I_G \} \mathrm{Pr} \{ \Phi_S | I_G \}}
  {\mathrm{Pr}\{ I_G \} \mathrm{Pr} \{ \Phi_S | I_G \} + \mathrm{Pr}\{ B_G \} \mathrm{Pr} \{ \Phi_S | B_G \}},
\end{equation}
\vspace{-.01cm}
\begin{equation}\vspace{-.05cm}
  \label{Pr_B_S}
  \mathrm{Pr} \{ B_S \}
  = \frac{\mathrm{Pr}\{ B_G \} \mathrm{Pr} \{ \Phi_S | B_G \}}
  {\mathrm{Pr}\{ I_G \} \mathrm{Pr}\{ \Phi_S | I_G \} + \mathrm{Pr}\{ B_G \} \mathrm{Pr} \{ \Phi_S | B_G \}}.
\end{equation}
$\mathrm{Pr} \{ \Phi_S | I_G \}$ and $\mathrm{Pr} \{ \Phi_S | B_G \}$ can be derived in a similar recursive manner as in (\ref{K_I_recursive}) and (\ref{K_B_recursive})
\vspace{-.1cm}
\begin{align} \vspace{-.1cm}
  \mathrm{Pr}\{ \Phi_S | I_G \}  &\!= \!
  \int_{t^P}^{+\infty}\!\!{ue^{-ut}dt} \!+\! 
  \mathrm{Pr}\{ \Phi_S | B_G \} \int_0^{t^P}{ue^{-ut}dt}, \label{Phi_S_I_G_recursive} \\
  \mathrm{Pr} \{ \Phi_S | B_G \}  &\!=\!
  \mathrm{Pr} \{ \Phi_S | I_G \}
  \int_0^{+\infty}{\int_s^{+\infty}{\lambda e^{-\lambda t}dtv e^{-vs}ds}}. \label{Phi_S_B_G_recursive}
\end{align}

Similar to (\ref{K_I_recursive}), the first term on the right side of (\ref{Phi_S_I_G_recursive}) corresponds to the successful packet reception on the first transmission attempt, and the second term corresponds to the case that the SD transmission is interrupted by the PD. In contrast, (\ref{Phi_S_B_G_recursive}) is different from (\ref{K_B_recursive}) in that $\mathrm{Pr}\{ \Phi_S | I_G \}$ should satisfy the condition that the packet isn't replaced by a newly-generated one during the BUSY period. By jointly considering (\ref{Phi_S_I_G_recursive}) and (\ref{Phi_S_B_G_recursive}), the $\mathrm{Pr} \{ \Phi_S | I_G \}$ and $\mathrm{Pr}\{ \Phi_S | B_G \}$ can be solved. 
\par
Since the generation process of packets and the state transition process of a channel are independent of each other, the probability of $I_G$ or $B_G$ is just the steady state probability of the IDLE or BUSY state, i.e. $\mathrm{Pr} \{ I_G \} = v/(u+v)$, $\mathrm{Pr} \{ B_G \} = u/(u+v)$. Consequently, $\mathrm{Pr} \{ I_S \}$ and $\mathrm{Pr} \{ B_S \}$ can be calculated by substituting the derived results to~\eqref{Pr_I_S} and~\eqref{Pr_B_S}.
\par
Compared to $\mathbb{E}[ S | I_G ]$ and $\mathbb{E}[ S | B_G]$, $\mathbb{E}[ S| I_S ]$ and $\mathbb{E}[ S| B_S ]$ are further conditioned on the event that the packet is finally successfully received at the SAP without being dropped. As a result, the two terms should further divide the conditional probabilities $\mathrm{Pr}\{ \Phi_S| I_G \}$ and $\mathrm{Pr}\{ \Phi_S | B_G \}$ on the basis of $\mathbb{E}[ S| I_G]$ and $\mathbb{E}[ S | B_G ]$, respectively. Based on (\ref{Phi_S_I_G_recursive}) and (\ref{Phi_S_B_G_recursive}), the expectation of $S$ conditioned on $I_G$ and $B_G$ are given as
\vspace{-.2cm}
\begin{align}
\label{S_I_G_recursive}
  \mathbb{E}[ S | I_S ] = {}& \frac{1}{\mathrm{Pr}\{ \Phi_S | I_G \}} \int_{t^P}^{+\infty}{t^P ue^{-ut} dt} \notag \\
  {}& + \frac{\mathrm{Pr} \{ \Phi_S | B_G \}}{\mathrm{Pr}\{ \Phi_S | I_G \}}
  \int_0^{t^P}{\left( t + \mathbb{E}[ S | B_S ] \right) ue^{-ut}dt}
\end{align}
\begin{align}
  \label{S_B_G_recursive}
  \!\!& \!\!\mathbb{E} \left[ S \left| B_S \right. \right] =\notag \\ 
  \!\!&\!\! \frac{\mathrm{Pr} \left\{ \Phi_S \left| I_G \right. \right\}}{\mathrm{Pr} \{ \Phi_S | B_G \}}
  \int_0^{+\infty}\!\!\!\!{\left( s + \mathbb{E} \left[ S \left| I_S \right. \right] \right)
  \int_s^{+\infty}\!\!\!\!{\lambda e^{-\lambda t}dtve^{-vs}ds}},
\end{align}
and the two terms can be solve jointly. Finally, $\mathbb{E}[S]$ can be expressed as follows:
\begin{equation}
  \label{S_final}
  \mathbb{E}[S] = 
  \frac{\left( l + \lambda t^P \right)e^{ut^P} - l + \frac{u}{u+v+\lambda}}
  {\lambda e^{ut^P} + v}.
\end{equation}
where $l \!=\! v/u \!+\! (u\!+\!v)/(u\!+\!v\!+\!\lambda)$.
The average AoI can now be characterized by substituting (\ref{Y2_2Y_final}) and (\ref{S_final}) into (\ref{AoIDefFinal}), given by
\begin{align}
  \label{AoI_final}
  \bar{\Delta} = {} & \frac{h^2 e^{2ut^P} + \left[ h(g-t^P) - \frac{1}{uv} \right]e^{ut^P} + \frac{g}{\lambda} + q}
  {he^{ut^P} + g} \notag \\
  {} & + \frac{\left( l + \lambda t^P \right)e^{ut^P} - l + \frac{u}{u+v+\lambda}}
  {\lambda e^{ut^P} + v},
\end{align}
and the energy efficiency can also be derived by substituting (\ref{K_final}), (\ref{Y_final}) and (\ref{TT}) into (\ref{EEdef}), given by
\begin{equation}
  \label{EE_final}
  \!EE \!= \!\frac{D}
  {\!\frac{P^T}{u}\!\left(\!e^{ut^P}\!\!-\!1\right)\! \!+\! \frac{P^C}{\lambda}\!+\! (P^S\!+\!P^C)\left(h e^{ut^P} \!+\! \frac{u}{v(u\!+\!v\!+\!\lambda)} \!-\! h\right)\!}.
\end{equation}
We further analyse the property of the derived average AoI and obtain the following lemma:
\begin{lemma}
\label{lemma_u}
For any finite ratio of average IDLE period and average BUSY period denoted by $k = v/u$, the average AoI tends to positive infinity when $u$ tends to $0$ or positive infinity.
\end{lemma}
\begin{IEEEproof}
See Appendix-B.
\end{IEEEproof}
According to Lemma~\ref{lemma_u}, for a fixed and moderate $k$, a very large $u$ corresponds to the case that both the continuous IDLE and BUSY period are statistically very short so that the transmission of the SD will be frequently interrupted by the PD. On the other hand, $u$ with an extremely small value corresponds to the case that both the continuous IDLE and BUSY period last very long in a statistical sense, and thus there will be a long time that the SD can obtain no transmission opportunity to update the status information of the monitored physical process. As a result, both cases will lead to a very high average AoI. We can also know from Lemma~\ref{lemma_u} that a large ratio $k$ doesn't guarantee a low average AoI, which seems counter-intuitive since a large $k$ usually means more transmission opportunities.

\section{Transmit Power Optimization}
In this section, the SD has to carefully decide its transmit power since a higher transmit power means a higher rate and lower AoI, however, this may cause lower energy efficiency. Since the status information will lose its value if it is outdated, the SD has a predefined AoI threshold denoted by $\Delta^{\mathrm{max}}$. Our objective is to maximize the energy efficiency of the SD subject to the average AoI constraint. The transmit power optimization problem is given by
\begin{subequations}
  \label{P1}
  \begin{alignat}{2}
    \mathbf{P1:}\max_{P^T}\quad & {EE( P^T )}&\quad & \tag{43}\\
    \text{s.t.}\quad & 0\leq P^T \leq P^{\mathrm{max}}, \label{P1:C1}\\
    &\bar{\Delta} ( P_n^T ) \leq {\Delta}^{\mathrm{max}}. \label{P1:C2}
  \end{alignat}
\end{subequations}
where $P^{\mathrm{max}}$ is the maximum transmit power. Constraint (\ref{P1:C1}) is the power budget for the SD. Constraint (\ref{P1:C2}) is used to guarantee that the average AoI of the SD should not exceed its AoI threshold. Due to the complexity of the energy efficiency and the average AoI derived in (\ref{EE_final}) and (\ref{AoI_final}), it is hard to directly analyse the convexity and monotonicity of the two terms w.r.t the transmit power $P^T$. Note that there is a one-to-one correspondence between the required transmission time of one packet $t^P$ and the transmit power $P^T$. More specifically, $t^P$ monotonously decreases with the increasing of $P^T$. As a result, we can change the optimization variable from $P^T$ to $t^P$ equivalently. Besides, according to the definition of the energy efficiency in (\ref{EEdef}), it is inversely proportional to the sum of three part of expected energy consumption between two consecutive successful receptions. Therefore, the optimization objective can be substituted by the sum of energy consumptions of transmi(ssion, spectrum sensing and static circuit, without changing the solution of the problem. With the above mentioned manipulations, we can obtain an equivalent problem to P1 given by
\vspace{-.1cm}
\begin{subequations}
  \label{P2}
  \begin{alignat}{2}
    \mathbf{P2:}\min_{t^P}\quad & {E^{\mathrm{sum}}( t^P )\triangleq  E^T( t^P )+ E^S( t^P ) + E^C( t^P )
    }&\quad & \tag{44}\\
    \text{s.t.}\quad & t^P \geq t^{\mathrm{min}}, \label{P2:C1}\\
    &\bar{\Delta} ( t^P ) \leq {\Delta}^{\mathrm{max}}. \label{P2:C2}
  \end{alignat}
\end{subequations}
\vspace{-.1cm}
where 
\vspace{-.1cm}
\begin{equation}\vspace{-.1cm}
  \label{min_t_P}
  t^{\mathrm{min}} = \frac{D}
  {B \log_2{\left( 1 + \frac{P^{\mathrm{max}}}{{N_0} B}\right)} }
\end{equation}
is the required packet transmission time at the maximum transmit power. To solve P2, we first analyse the monotonicity and the convexity of $E^{\mathrm{sum}}( t^P )$ and $\bar{\Delta} ( t^P )$ and obtain the following two lemmas.

\begin{lemma}
\label{lemma_E_sum}
$E^{\mathrm{sum}}(t^P)$ is convex in $t^P \in (0,\sqrt{\frac{D\ln2}{B u}})$ and strictly monotonously increases for $t^P \in [\sqrt{\frac{D\ln2}{B u}},+\infty)$. Besides, $E^{\mathrm{sum}}(t^P)$ is a quasi-convex function and first decreases and then increases in interval $(0,+\infty)$.
\end{lemma}
\begin{IEEEproof}
  See Appendix-C.
\end{IEEEproof}

\begin{lemma}
  \label{lemma_AoI}
  $\bar{\Delta} ( t^P )$ is a strictly monotonously increasing function in interval $(0,+\infty)$.
\end{lemma}
\begin{IEEEproof}
  See Appendix-D.
\end{IEEEproof}
With the convexity and monotonicity in Lemma~\ref{lemma_E_sum} and Lemma~\ref{lemma_AoI}, P2 and its equivalent problem P1 can be solved using binary search and gradient decent method, which is elaborated in Algorithm~1.
\begin{algorithm}[!t]
\captionsetup{font={small}}
\caption{TPOA: Transmit Power Optimization Algorithm}
\label{alg:TPOA}
\renewcommand{\algorithmicrequire}{\textbf{Input:}}
\renewcommand{\algorithmicensure}{\textbf{Output:}}
\begin{small}
\begin{algorithmic}[1]
\linespread{1}\selectfont
\Require $\lambda,D,{N_0},{\Delta}^{\mathrm{max}},P^{\mathrm{max}},u,v,B$
\Ensure solution to P2
\State Calculate $t^{\mathrm{min}}$ and $\bar{\Delta}(t^{\mathrm{min}})$ from (\ref{min_t_P}) and (\ref{AoI_final}).
\If {$\bar{\Delta}(t^{\mathrm{min}}) > {\Delta}^{\mathrm{max}}$}
\parState {P2 has no feasible solution. The SD isn't able to meet its information freshness requirement.}
\Else
\State Select some large $t^{\prime}$ that satisfies $\bar{\Delta}(t^{\prime})>{\Delta}^{\mathrm{max}}$.
\parState {Do binary search in interval $[t^{\mathrm{min}},t^{\prime}]$ until find a $t^{\mathrm{max}}$ that satisfies $\bar{\Delta}(t^{\mathrm{max}}) = \Delta^{\mathrm{max}}.$}
\State Calculate $\Omega = (0,\sqrt{\frac{D\ln2}{B u}})\cap[t^{\mathrm{min}},t^{\mathrm{max}}]$.
\If ${\Omega = \emptyset}$
\State ${t^P}^{\ast} = \sqrt{\frac{D\ln2}{B u}}$.
\Else
\State Obtain ${t^P}^{\ast}$ in $\Phi$ using gradient decent method.
\EndIf
\State Calculate ${P^T}^{\ast}=P^T({t^P}^{\ast})$, $EE^{\ast}=EE({P_n^T}^{\ast})$.
\EndIf
\end{algorithmic}
\end{small}
\end{algorithm}
\section{Numerical Simulations}
In this section, we present the numerical results of the energy efficiency and AoI performance of the considered CR-IoT system. In the simulations, we set the bandwidth of the licensed channel to be $B=180$ kHz, which equals the bandwidth of a sub-channel in NB-IoT, and set the spectrum sensing power, the static circuit power and the maximum transmit power to be $P^S = 1\times10^{-3}$ W, $P^C = 1\times10^{-4}$ W and $P^{\mathrm{max}}=0.1$ W, respectively. The equivalent noise power per unit bandwidth at the SAP in the simulation is set to be ${N_0}=N_0^R/\eta$, where $N_0^R$ is the noise power per unit bandwidth at the receiver of SAP and $\eta$ is the path loss of SD. We set $N_0^R=-110$ dBm and adopt $\eta = L^{-\theta}$ for the simulation of path loss with path loss factor $\theta=3$ and the distance between the SD and the SAP denoted by $L$.
\par
\begin{figure}[t]
  \captionsetup{font={small}}
  \centering
  \subfloat[Energy efficiency.]{
  \label{fig_ee_u}
  \includegraphics[width=0.75\linewidth, trim = 30 5 20 30]{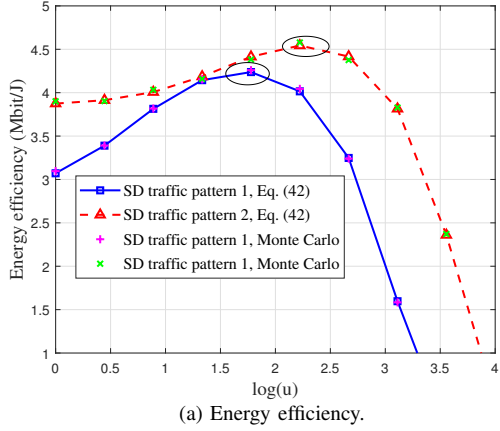}
  }
  \hfill
  \subfloat[Average AoI.]{
  \label{fig_aoi_u}
  \includegraphics[width=0.75\linewidth, trim = 30 5 20 20]{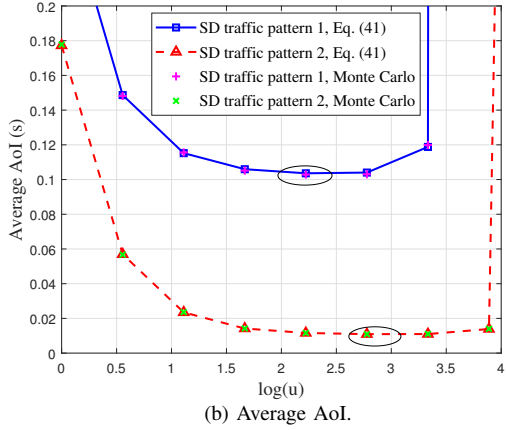}
  }
  \caption{{Energy efficiency and AoI performance versus CTMC parameter $u$ where $L=250$. The two SD traffic patterns are $D_1 = 400, \lambda_1 = 10$ and $D_2=100,\lambda_2=100$.}}
 \vspace{-.53cm}
\end{figure}

We first analyse the energy efficiency and AoI performance under different traffic patterns in Fig. $2$. To characterize different PD traffic patterns, we fix the IDLE/BUSY ratio while varying the parameter $u$. It can be seen that the results derived in (\ref{EE_final}) and (\ref{AoI_final}) coincide well with the Monte Carlo simulation results, which verifies the characterization on the energy efficiency and the average AoI. Besides, as shown in Fig. \ref{fig_ee_u}, the energy efficiency under both the two traffic patterns first increases then decreases with the increasing $u$, which means there exists a best $u$ to achieve the highest energy efficiency. Similarly, in Fig. \ref{fig_aoi_u}, the average AoI first decreases then increases and tends to positive infinity on both sides, which verifies Lemma~\ref{lemma_u}. More importantly, the $u$ that maximizes the energy efficiency or minimize the average AoI for the two SD traffic patterns are different, which implies that the SDs with different traffic have different preferences toward different PD traffic patterns. Moreover, comparing the two subfigures, for a single SD, the best $u$ w.r.t the two performance indcators are also different, which indicates that the SD have different preferences towards the PD traffic pattern for the two performance demands.

\begin{figure}[t]
  \captionsetup{font={small}}
  \centering
  \subfloat[Energy efficiency.]{
  \includegraphics[width=0.75\linewidth,  trim = 30 5 20 30]{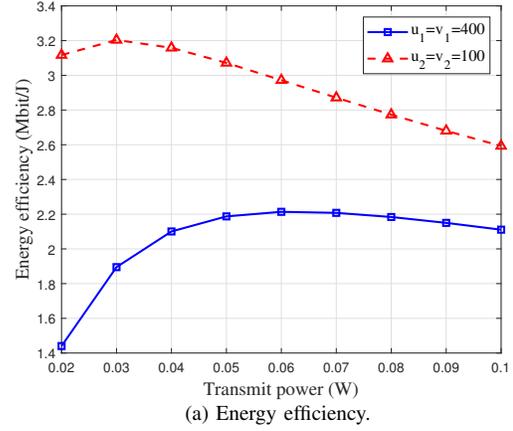}
  \label{fig_ee_transmitpower}}
  \hfil
  \subfloat[Average AoI.]{
  \includegraphics[width=0.75\linewidth,  trim = 30 5 20 20]{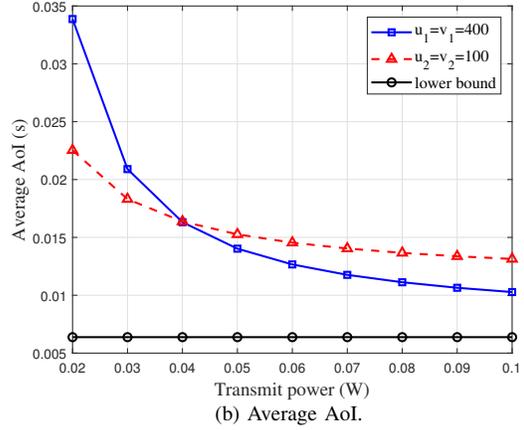}
  \label{fig_aoi_transmitpower}}
  \caption{{Energy efficiency and AoI performance versus transmit power where $L=300$, $D = 400$ and $\lambda = 200$.}}
    \vspace{-.53cm}
    % \vspace{-.53cm}
\end{figure}

Next, we present the energy efficiency and AoI performance of a SD under different transmit power in Fig. \ref{fig_ee_transmitpower} and \ref{fig_aoi_transmitpower}. The solid and dashed lines in the two figures correspond to two PD traffic patterns, i.e. different CTMC parameters $u$ and $v$, under the same IDLE/BUSY ratio $k=v/u$ . The black line in Fig. \ref{fig_aoi_transmitpower} represents the sum of the average packet generation interval and the required transmission time for a single packet, which is a lower bound of the average AoI. It can be observed that with the increasing transmit power, the energy efficiency of the SD first increases and then decreases, while the average AoI keeps decreasing, which verifies Lemma~\ref{lemma_E_sum} and Lemma~\ref{lemma_AoI}. We can also find that although the energy efficiency under PD traffic pattern~$1$ is lower than $2$, the lowest reachable average AoI under pattern~$1$ is lower than that under pattern~$2$. As a result, when the average AoI constraint is strict, e.g. 2 times of the lower bound, pattern~$2$ isn't able to satisfy the information freshness requirement of the SD while pattern~$1$ is, so the SD would prefer pattern~$1$ with a lower energy efficiency. This implies that the transmit power control introduces a trade-off between the energy efficiency and the average AoI.

\section{Conclusion}
In this paper, we have investigated the impact of heterogeneous traffic pattern on the energy efficiency and average AoI of the SD in a CR-IoT system. The closed-form expressions of the energy efficiency and the average AoI have been derived. We have designed for the SD an optimal transmit power optimization algorithm aiming at maximizing the energy efficiency while satisfying the average AoI constraint. 
% To characterize and solve the problem, we rigorously proved that the average AoI is convex in the transmission time and that the energy consumption is convex and monotonically increasing in the transmission time. 
The solution is facilitated by exploring the convexity and monotonicity of the objective and constraint functions. % we have designed for the SD 
The  numerical results   have confirmed our analytical model. In addition, we  showed that the SD has different preferences toward different PD traffic patterns, and there is a tradeoff between the energy efficiency and the average AoI.

% \section*{Acknowledgment}

% The preferred spelling of the word ``acknowledgment'' in America is without 
% an ``e'' after the ``g''. Avoid the stilted expression ``one of us (R. B. 
% G.) thanks $\ldots$''. Instead, try ``R. B. G. thanks$\ldots$''. Put sponsor 
% acknowledgments in the unnumbered footnote on the first page.

\section*{Appendix} 
\subsection{Proof of Lemma 2}
When $u$ tends to zero, for a finite $k=v/u$, substitute $k$ for $v$ in (\ref{Y2_2Y_final}) and (\ref{S_final}), and utilize the Taylor expansion of term $\mathrm{exp}({ut^P})$ and $\mathrm{exp}({2ut^P})$, the limit of term $\mathbb{E}[Y^2]/2\mathbb{E}[Y]$ and $\mathbb{E}[S]$ can be expressed as\vspace{-.1cm}
\begin{align}
  \lim_{u\rightarrow 0}{\frac{\mathbb{E}[Y^2]}{2\mathbb{E}[Y]}} &= 
  \lim_{u\rightarrow 0}{\frac{\frac{1}{k^2}\left(t^P+\frac{1}{\lambda}\right) \frac{1}{u} + o\left(\frac{1}{u}\right)}{\left(1+\frac{1}{k}\right)\left( t^P + \frac{1}{\lambda} \right) + o(1)}} =
  +\infty, \\
  \lim_{u\rightarrow 0}{\mathbb{E}[S]} &= 
  \lim_{u\rightarrow 0}{\frac{\lambda t^P + o(1)}{\lambda + o(1)}} = 
  t^P.
\end{align} 
Therefore, the limit of average AoI at $u \rightarrow 0$ is given by % tends to zero is given by
\vspace{-.1cm}
\begin{equation}\vspace{-.1cm}
  \lim_{u\rightarrow 0}{\bar{\Delta}} = 
  \lim_{u\rightarrow 0}{\left(\frac{\mathbb{E}[Y^2]}{2\mathbb{E}[Y]} + \mathbb{E}[S] \right)} =
  +\infty.
\end{equation}
\par
When $u$ tends to positive infinity, as for $\mathbb{E}[Y^2]/2\mathbb{E}[Y]$, the $\mathrm{exp}({2ut^P})$ term in the numerator and the $\mathrm{exp}({ut^P})$ term in the denominator play a dominant role, and we can obtain that\vspace{-.1cm}
\begin{equation}\vspace{-.1cm}
  \lim_{u\rightarrow +\infty}{\frac{\mathbb{E}[Y^2]}{2\mathbb{E}[Y]}} = 
  \lim_{u\rightarrow +\infty}{\frac{{\left(1+\frac{1}{k}\right)}^2 \frac{1}{u^2} e^{2ut^P}}{\left(1+\frac{1}{k}\right) \frac{1}{u} e^{ut^P}}} =
  +\infty.
\end{equation}
Similarly, the $\mathrm{exp}({ut^P})$ terms play a dominant role in both the numerator and denominator of the term $\mathbb{E}[S]$ when $u$ tends to positive infinity, and the limit is given by\vspace{-.1cm}
\begin{equation}\vspace{-.1cm}
  \lim_{u\rightarrow +\infty}{\mathbb{E}[S]} = 
  \lim_{u\rightarrow +\infty}{\frac{\left(k+\frac{u+ku}{u+ku+\lambda}+\lambda t^P\right)e^{ut^P}}{\lambda e^{ut^P}}} = 
  \frac{1+k+t^P}{\lambda}.
\end{equation}
Finally we can get\vspace{-.1cm}
\begin{equation}\vspace{-.1cm}
  \lim_{u\rightarrow +\infty}{\bar{\Delta}} = 
  \lim_{u\rightarrow +\infty}{\left(\frac{\mathbb{E}[Y^2]}{2\mathbb{E}[Y]} + \mathbb{E}[S] \right)} =
  +\infty,
\end{equation}
and Lemma~\ref{lemma_u} is proved.  

\subsection{Proof of Lemma 3}
According to (\ref{K_final}) and (\ref{Y_final}), we have %$E^{S}( t^P )$ and $E^{C}( t^P )$ can be expressed as
\vspace{-.1cm}
\begin{align}
  E^{S}( t^P ) & = P^S \left( he^{ut^P} + \frac{u}{v(u+v+\lambda)} - h \right), \label{ES} \\ 
  E^{C}( t^P ) & = P^C \left( he^{ut^P} + \frac{1}{\lambda} + \frac{u}{v(u+v+\lambda)} - h  \right). \label{EC}
\end{align}
From (\ref{ES}) and (\ref{EC}), we know that both $E^{S}( t^P )$ and $E^{C}( t^P )$ are convex and strictly monotonously increasing w.r.t $t^P$ in interval $(0,+\infty)$. As for the convexity and monotonicity of $E^{S}( t^P )$, its expression can be obtained from (\ref{EEdef}) and (\ref{TT}), which is given by\vspace{-.1cm}
\begin{equation}\vspace{-.1cm}
  \label{ET}
  E^{T}( t^P ) = N_0 B \left[ \mathrm{exp}\left( \frac{D \ln{2}}{Bt^P} \right) -1 \right]
  \left[ \mathrm{exp}\left( ut^P \right) -1 \right].
\end{equation}
According to the structure of $E^{S}( t^P )$ in (\ref{ET}), consider a function of this kind\vspace{-.1cm}
\begin{equation}\vspace{-.1cm}
  \label{f(x)}
  f(x)= \left( e^{ax} -1 \right) \left( e^{\frac{b}{x}} -1  \right), a>0,b>0,x\in(0,+\infty).
\end{equation}
Take the first derivative of $f(x)$ and we can get\vspace{-.1cm}

\begin{equation}\vspace{-.1cm}
  \label{f'(x)}
  f^{\prime}(x) = e^{ax+ \frac{b}{x}}\left( a - \frac{b}{x^2} -ae^{-\frac{b}{x}} + \frac{b}{x^2} e^{-ax} \right).
  \end{equation}
Let $s(x) =  a - \frac{b}{x^2} -ae^{-\frac{b}{x}} + \frac{b}{x^2} e^{-ax}$, we can easily find that $s(\sqrt{\frac{b}{a}})=0$ and $f^{\prime}(\sqrt{\frac{b}{a}})=0$. When $x>\sqrt{\frac{b}{a}}$, we can obtain that the derivative of $s(x)$
\vspace{-.1cm}
\begin{equation}\vspace{-.1cm}
  \label{s'(x)}
  s^{\prime}(x) 
  =  \frac{2b}{x^2} \left( 1 - e^{-ax} \right) + \frac{ab}{x^2} \left( e^{-\frac{b}{x}} - e^{-ax} \right) >0.
\end{equation}
Therefore, $f^{\prime}(x)>0$ when $x>\sqrt{\frac{b}{a}}$, which means $f(x)$ is strictly monotonously increasing in interval $[\sqrt{\frac{b}{a}},+\infty)$.

When $x\in(0,\sqrt{\frac{b}{a}})$, take the second-order derivative of $f(x)$

%\begin{align}
%  \label{f''(x)}
%  f^{\prime\prime}(x) 
%  = & {\left( a - \frac{b}{x^2} \right)}^2 e^{ax+\frac{b}{x}} + \frac{2b}{x^3} e^{ax+\frac{b}{x}} - a^2e^{ax} - \frac{2b}{x^3} e^{\frac{b}{x}} - \frac{b^2}{x^4} e^{\frac{b}{x}} \notag \\
%  = & \left( e^{ax} -1 \right) e^{\frac{b}{x}} {\left( a - \frac{b}{x^2} \right)}^2 + a^2\left( e^{\frac{b}{x}} - e^{ax} \right) \notag \\ 
%  &+ \frac{2be^{\frac{b}{x}}}{x^3} \left( e^{ax} - ax - 1 \right)
%  \notag \\
%  > & 0.
%\end{align}
 
$\!\!\!\!\begin{array}{ll}
  \!\!\!\!f^{\prime\prime}(x) \!\!
   &\!\!\!\!=\! {\left( a \!-\! \frac{b}{x^2} \right)}^2 e^{ax+\frac{b}{x}} \!+\! \frac{2b}{x^3} e^{ax+\frac{b}{x}} - a^2e^{ax} \!-\! \frac{2b}{x^3} e^{\frac{b}{x}} - \frac{b^2}{x^4} e^{\frac{b}{x}} \notag \\
  & \!\!\!\!=\! \left( e^{ax} \!\!-\!1 \!\right) e^{\frac{b}{x}} {\left( a \!-\! \frac{b}{x^2}\! \right)}^{\!2} \!\!+\! a^2\left( e^{\frac{b}{x}} \!-\! e^{ax} \! \right)  \!+\! \frac{2be^{\frac{b}{x}}}{x^3} \!\left( e^{ax} \!\!-\! ax \!-\! 1 \right)
  \notag \\
  &\!\!\!\!>\!  0.
\end{array}$

Therefore, $f(x)$ is convex in interval $(0,{\textstyle {\sqrt{\frac{b}{a}}}})$. Let $a=u$ and $b=\frac{D\ln2}{Bu}$, we find that $E^T(t^P)$ is  convex  in $t^P \in (0,\sqrt{\frac{D\ln2}{Bu}})$ and strictly monotonously   in $t^P \in[ \sqrt{\frac{D\ln2}{Bu}}, +\infty)$. Along with the convexity and monotonicity of $E^{S}( t^P )$ and $E^{C}( t^P )$, we come to the conclusion that $E^{\mathrm{sum}}( t^P )$ is a convex function in $(0,\sqrt{\frac{D\ln2}{Bu}})$ and strictly monotonously increases in interval $[ \sqrt{\frac{D\ln2}{Bu}}, +\infty)$ and Lemma~\ref{lemma_E_sum} is proved.

\subsection{Proof of Lemma 4}
To analyse the monotonicity of $\mathbb{E}[S]$, we take the derivative of it in (\ref{S_final}) w.r.t $t^P$ and obtain that
\vspace{-.1cm}
\begin{equation}\vspace{-.1cm}
  \label{dE[S]}
  \frac{\mathrm{d}\mathbb{E}[S]}{\mathrm{d}t^P} = 
  \frac
  {e^{ut^P} \left[ {\lambda}^2 e^{ut^P} + v(\lambda + \lambda u t^P + v + 1 + \frac{u(u+\lambda+1)}{u+v+\lambda} )
  \right]}
  {( \lambda e^{ut^P} + v )^2} > 0
\end{equation}

Similarly, we take the derivative of $\mathbb{E}[Y^2]/2\mathbb{E}[Y]$ as follows,
\vspace{-.1cm}
\begin{equation}\vspace{-.1cm}
  \label{dEY22EY}
  \frac{\mathrm{d}}{\mathrm{d}t^P} \frac{\mathbb{E}[Y^2]}{2\mathbb{E}[Y]} = 
  \frac{e^{ut^P}r(t^P)}{{\left( he^{ut^P} + g\right)}^2},
\end{equation}
where
\vspace{-.1cm}
  \begin{align}
  \label{r(t)}
  r(t^P) =& uh^2e^{2ut^P} + (2ug-1)h^2e^{ut^P} - uhgt^P \notag \\ 
  &+ uhg^2 - hg(1+\frac{u}{v}) - \frac{h+g}{v} + \frac{u(v+\lambda)h}{v(u+v+\lambda)^2}.
\end{align}
We can easily find that 
\vspace{-.1cm}
  \begin{equation}\vspace{-.1cm}
  \label{r(0)}
  r(0) = \frac{u(u+v)^2(2v+\lambda)}{\lambda v^2 (u+v+\lambda)^2} > 0.
\end{equation}
We further take the derivative of $r(t^P)$ and obtain that
\vspace{-.1cm}
\begin{equation}\vspace{-.1cm}
  \label{dr(t)}
  \frac{\mathrm{d}r(t^P)}{\mathrm{d}t^P} 
  = uh\left(he^{ut^P}+g\right) \left[ 2\left(1+\frac{u}{v}\right)e^{ut^P} - 1\right] > 0.
\end{equation}
As a result, $r(t)>0$ for $t^P\in(0,+\infty)$ and thus $\mathbb{E}[Y^2]/2\mathbb{E}[Y]$ strictly monotonously increases $(0,+\infty)$. Lemma~\ref{lemma_AoI} is proved.

\small
\bibliographystyle{IEEEtran}
%\linespread{1.15}\selectfont %调节参考文献间距
\bibliography{IEEEabrv,Ref.bib}

\end{document}